\documentclass[a4paper,12pt]{article}

\usepackage{color}
\usepackage{amsmath}
\usepackage{amssymb}
\usepackage{graphicx}
\usepackage{hyperref}
\usepackage{subfigure}

\def \OS{\texttt{OS}}
\def \FL{\texttt{FL}}
\def \MA{$\mathrm{M_A}$}
\def \MC{$\mathrm{M_C}$}
\def \eg{{\em e.g.}}

\def \gev{{\rm GeV}}
\def \tev{{\rm TeV}}
\def \GA{$\mathrm{\Gamma_A}$}
\def \GC{$\mathrm{\Gamma_C}$}

\begin{document}

\begin{center} 

{\large\bf Dijet resonances, widths and all that }\\[3ex]

Debajyoti Choudhury$^a$, Rohini M. Godbole$^{b}$ and Pratishruti Saha$^a$\\

\vspace*{5pt}

\begin{footnotesize}
$^a$ Department of Physics and Astrophysics, University of Delhi, 
Delhi 110 007, India.\\
$^b$ Centre for High Energy Physics, Indian Institute of Science, 
Bangalore, 560 012, India.\\
\end{footnotesize}

\end{center} 

\vspace*{40pt}
 
\begin{abstract} 
\noindent
The search for heavy resonances in the dijet channel is part of the
on-going physics programme, both at the Tevatron and at the LHC.
Lower limits have been placed on the masses of dijet resonances
predicted in a wide variety of models. However, across experiments,
the search strategy assumes that the effect of the new particles is
well-approximated by on-shell production and subsequent decay into a
pair of jets.  We examine the impact of off-shell effects on such
searches, particularly for strongly interacting resonances.

\vspace*{40pt}
\noindent
\texttt{PACS Nos:} 14.70.Pw, 12.38.Qk\\ 
\texttt{Key Words:} dijet, finite-width, Tevatron, LHC
\end{abstract}

\section{Introduction}
Dijet production is an integral part of studies conducted at any
hadron collider. Apart from being important for the ratification of
our understanding of QCD, this process also provides a fertile ground
for new physics searches, particularly searches for new particles that
may appear as resonances in the dijet invariant mass spectrum.
The large production cross-sections mean that
significant conclusions may be drawn even with low integrated
luminosities. Of course, QCD itself gives rise to a large
background which is a challenge that any new physics search has to
contend with. However, in case the new physics signal appears in the
form of a localised excess in production rates in a particular region
of phase space ({\eg} a resonance), it is often possible to subdue the
background with appropriate kinematic cuts.

While no new physics beyond the 
Standard Model (SM) has been discovered yet,
experimental collaborations at both the Tevatron and the LHC 
have been using the dijet channel to
place limits on masses of new particles in a wide variety of 
models~\cite{CDF_dijet_2008,CMS_dijet_2011,ATLAS_dijet_2011}.  The
strategy has been to compute upper bounds on $\sigma \cdot {\cal B} \cdot {\cal
  A}$ (where $\sigma$ is the cross-section for on-shell production of
the particle, ${\cal B}$ is the branching fraction into a pair of jets
and ${\cal A}$ is the experimental acceptance) and hence rule out
certain mass ranges for the new particles. In essence, $\sigma \cdot
{\cal B}$ is the excess that arises due to the presence of the new
particle.  This implies the assumption that the new particle under
consideration has a small width. 
However, for some strongly decaying particles, which the Tevatron/LHC
have excluded upto a large mass, this assumption may not hold true.
Examples are
particles such as axigluons and colorons which typically have a width
\begin{small}$\gtrsim$\end{small} $10\%$ of their mass and, are
certainly not `narrow resonances'. However, in
the context of a hadron collider, it is also important to remember
that the `observed width' as would be measured from the invariant mass
distribution is not the true decay width of the particle associated
with the resonance. The shape of the resonance gets distorted by
fluxes of the initial state particles which, in turn, depend
on the parton level center-of-mass energy.
Further, experimental reconstruction of jets is a challenging task.
Limitations of detector resolution and reconstruction algorithms 
affect the resolution of the jet-jet invariant mass $m_{jj}$.
If the aforementioned effects 
overwhelmingly dominate the observed width, then 
the theoretical approximation of
`narrow-width' may be inconsequential. In this note, we examine such
aspects of dijet resonance searches in an attempt to compare the
relative merits of different search strategies.

The rest of this article is organised as follows. We begin 
by briefly recapitulating, in Section \ref{sec:models}, a few model templates 
with particular emphasis on their status {\em vis \`{a} vis}  
searches at the Tevatron and the LHC. 
Section \ref{sec:numerical} details our calculations and 
delineates the numerical effects due to the off-shell contributions. 
Finally, we conclude in Section \ref{sec:summary}.

\section{Dijets and new physics}
\label{sec:models}
With both the Tevatron and the LHC accumulating substantial
luminosity, several studies have looked at possible
resonances. While resonances decaying into leptons (such as a
hypothetical $Z'$) are relatively easy to look for, of particular
interest are particles that preferentially decay into hadronic
states. Indeed, considerable interest has been generated by two
Tevatron measurements, viz.  the reported excess~\cite{wjj_expt} in
the $W j j $ final state on the one hand, and the longstanding
discrepancy in the forward-backward asymmetry in $t \bar t$
production~\cite{ttb_expt} on the other. While several models have
been proposed as solutions to these deviations from the SM, it is
imperative that they be examined {\em vis \`{a} vis} other
processes. With each such explanation positing new couplings
involving quarks, dijet production is a natural theatre for such
investigations~\cite{Bai:2011ed}.
Indeed, a recent study~\cite{Han:2010rf} has
attempted a model independent study of colored resonances.

The simplest algorithms for new particle searches naturally
concentrate on situations wherein the role of the said particle, in any
process involving only SM particles as asymptotic states, can be
well-approximated by a narrow resonance. Unfortunately though, this
approximation is often rendered invalid.  
Further,
under certain conditions, even the assumption of a constant width
 may need correction.  Numerous examples abound in low-energy hadron
physics (see for example, Ref.~\cite{Frazer:1960zzb}). Indeed, this
effect turned out to be significant even for a narrow resonance such as
the $Z$~\cite{Bardin:1988xt}. More recently, this question has been
examined in the context of production of
top-pairs~\cite{Choudhury:2007ux,Frederix:2007gi},
dijets~\cite{Haisch:2011up}, a heavy SM Higgs
boson~\cite{Choudhury:2002qb} and hypothetical $W'$s~\cite{Wpr_width}
and $Z'$s~\footnote{Although  particles such as a heavy SM Higgs, $W'$
  or $Z'$ do not decay strongly, their large masses \mbox{($\gtrsim 600 \, \gev$)}
  still result in a substantially large width. Of particular importance 
  is the fact that, for such widths, the interference with non-resonant 
  contributions to the amplitude become important, thereby rendering 
  the on-shell approximation rather inaccurate.}~\cite{Godbole:2010kr}.

In this study, we consider axigluons\footnote{Axigluons are of
    particular interest in the context of the forward-backward
    asymmetry in $t \bar t$ production. See, for example, 
    Refs.~\cite{Choudhury:2007ux,Frampton:2009rk,Choudhury:2010cd}.}
and colorons as templates for broad dijet resonances. These are color octet
gauge bosons appearing in certain classes of models that hypothesize
an extended color gauge group $SU(3)_A \otimes SU(3)_B$ at high
energies. The extended gauge symmetry is broken spontaneously to
$SU(3)_C$ that one associates with strong interactions in the SM. 
Axigluons and colorons correspond to the broken generators in
the respective models and, hence, are massive.
The coupling of axigluons to quarks is given by $g_s
\gamma_{\mu} \gamma_5$ while that of colorons is $g_s \gamma_{\mu}
\cot\xi$, $\xi$ being the angle that characterizes the mixing between
the two $SU(3)$ groups.  In certain variants of the model, the coloron
couples preferentially to the third generation quarks. However, here
we consider only the {\em flavor-universal} coloron, for which
$\cot\xi \lesssim$ 4 in order that the model remains in its Higgs
phase. Details of
axigluon and coloron models can be found in
Refs.~\cite{axigluon,limits_axigluon} and
Refs.~\cite{coloron,limits_coloron} respectively. Over the years,
limits on axigluon and coloron mass 
(denoted henceforth by \MA and \MC, respectively)
have been upgraded continuously
based on experiments that have been in operation at various
times~\cite{dijet_searches}.  The current experimental limits are due
to searches in the dijet channel at the LHC. The ATLAS experiment has
ruled out \MA $<$ 3.32 TeV~\cite{ATLAS_dijet_2011} while CMS has ruled
out axigluons and colorons of mass below 2.47
TeV~\cite{CMS_dijet_2011}. Before this, the CDF experiment at
the Tevatron had placed a lower limit of 1250 GeV on \MA,
\MC~\cite{CDF_dijet_2008}. However, barring the analysis of
  Ref.~\cite{Choudhury:2007ux} (included by the Particle Data
  Group~\cite{pdg}), the rest implicitly incorporate the narrow-width 
approximation, and thus need to be accepted with care.

\section{Numerical Analysis}
\label{sec:numerical}
In this analysis, for the Tevatron as well as the LHC, 
we consider representative masses of the aforementioned particles 
in the range that is well within the kinematical reach of the 
respective machines. Then, for each such representative case, 
we compute the signal in two ways --
first, by considering {\em on-shell} production and subsequent decay 
of the new particle, and second, by computing the {\em full} cross-section 
including off-shell effects and all contributing amplitudes 
($s$-, $t$- and $u$- channel etc.). We label these two cases by 
{\OS} and {\FL} respectively.

The computations are performed using CalcHEP~\cite{CalcHEP}.
The CTEQ6L parton distributions~\cite{CTEQ} are used along 
with the compatible value
for $\alpha_s$ as obtained using the 2-loop 
$\beta$-function\footnote{While this might seem paradoxical given that 
we are computing only the leading-order matrix elements, note that 
the CTEQ collaboration uses $\alpha_s$ calculated at NLO to extract the 
CTEQ6L distributions. Furthermore, the use of CTEQ6L1 distributions alongwith 
$\alpha_s({\rm LO})$ does not qualitatively change our conclusions.}.
The renormalization and factorization scales are set to the sub-process 
center of mass energy.

\subsection{At the Tevatron}
As a representative example close to the lower bound achieved by CDF,
we consider an axigluon of mass \MA = 1.2 TeV. 
The natural width of this particle is 100 GeV.
To enhance the signal
to background (SM) ratio, we only generate events where the jets have
a $p_T $ of at least\footnote{This is nothing but the imposition of
  $\mbox{\sc ckin}(3) > 150$ GeV in {\sc Pythia}~\cite{pythia}.} 150 GeV.
Since we are primarily interested in a localised excess in the
$m_{jj}$ distribution, we must effect a fit to the $m_{jj}$ spectrum
significantly away from the excess. An excellent
fit is obtained~\cite{CDF_dijet_2008} in terms of the
four-parameter function $f(x) = a_0 \, (1 - x)^{a_1} \, x^{a_2 + a_3
\, \ln x} \, $ where $x \equiv m_{jj} / \sqrt{s_{pp}}$. 
Interestingly, a Gaussian in $x$ gives a (three-parameter) fit that 
is virtually as good (in terms of $\chi^2$ per degree of freedom).

\begin{figure}[!htbp]
\hspace*{-20pt}
\centering
\subfigure[]
{
\includegraphics[width=3.5in,height=3.0in]{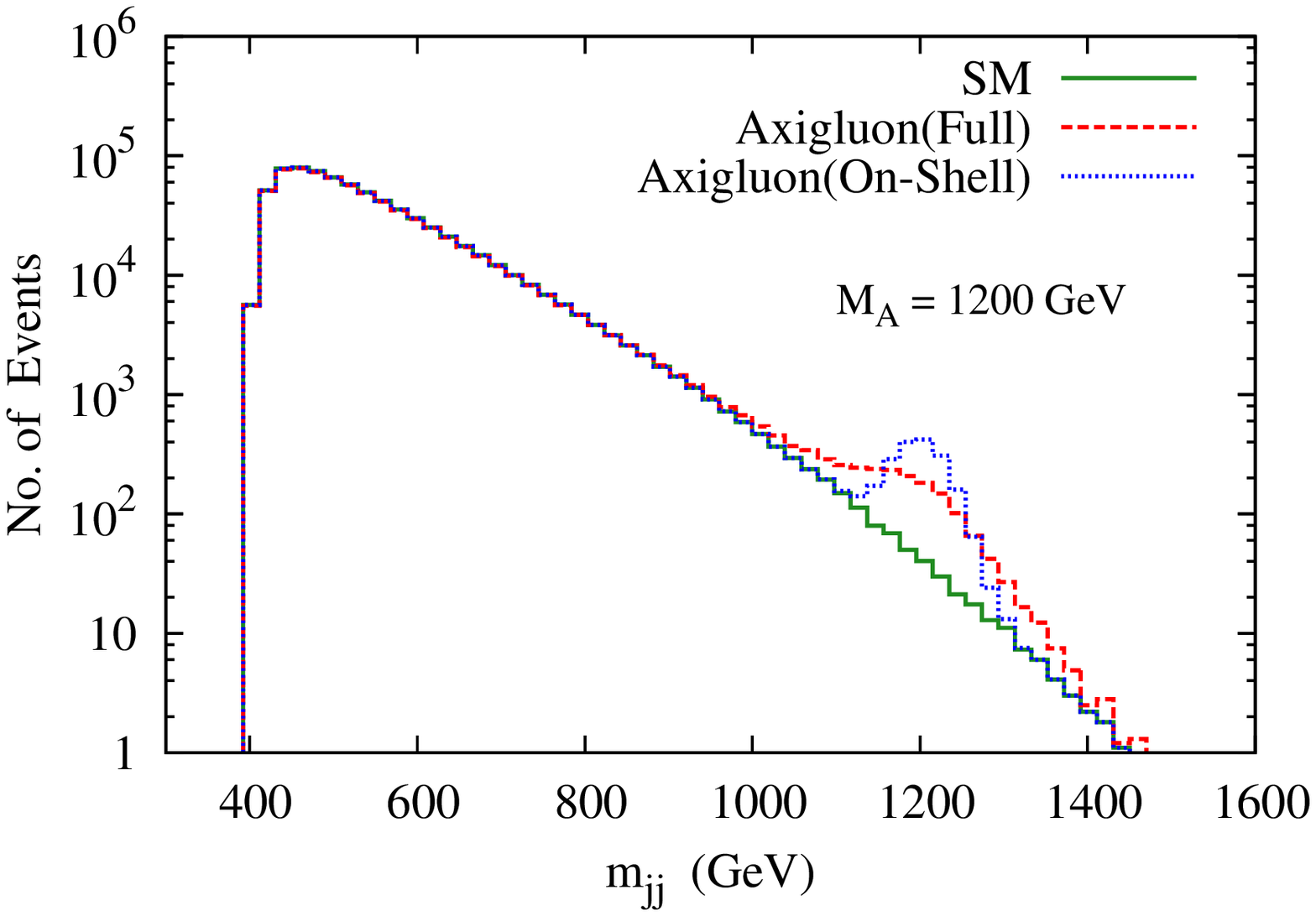}
\label{fig:TeV_Axi_pT200_smeared}
}\hspace*{-10pt}
\subfigure[]
{
\includegraphics[width=3.5in,height=3.0in]{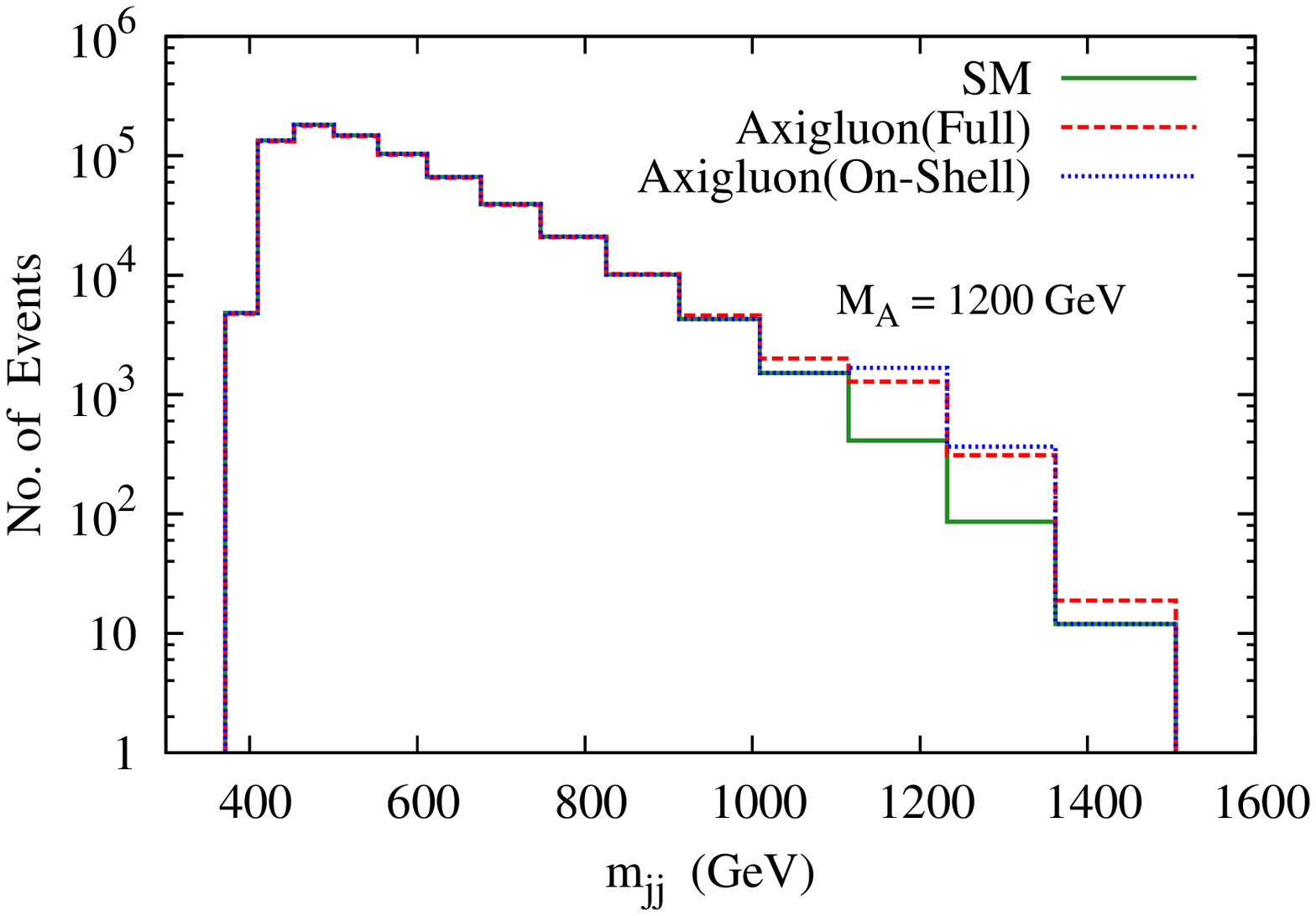}
\label{fig:TeV_Axi_pT200_rebinned_smeared}
}
\vspace*{-10pt}
\caption{\em The dijet invariant mass spectrum at the Tevatron corresponding 
to \MA = \mbox{$1200$ \gev} (the corresponding width is \GA = $100$ {\gev}).
An integrated luminosity of $10$ fb$^{-1}$ has been
assumed and a restriction of $p_T > 200$ {\gev} imposed on each of the 
two highest $p_T$ jets. The left panel corresponds to a constant 
$m_{jj}$ bin width, while the right panel redistributes the events in bins
with $\delta m_{jj} = m_{jj}^{\rm av}/10$.}
\label{fig:TeV_Axi_pT200}
\end{figure}

Once the aforementioned (theoretical) SM spectrum is obtained, the
invariant mass spectrum for both the {\OS} and {\FL} case can, then,
be compared with it.  In the infinite resolution limit, the {\OS}
distribution would be characterised by a sharp spike in the bin
corresponding to $m_{jj}$ = 1.2 TeV. However, in an experimental
situation, the spike gets smeared due to detector resolution effects.
Bearing this in mind, we apply a Gaussian smearing to the energy of
all the final state particles with $\delta E_T/E_T = 50\%/\sqrt{E_T
  ({\rm GeV})} \oplus 3\%$ (this being the resolution of the central
hadron calorimeter for CDF~\cite{CDF_dijet_2008}).  To account for
possible upward scaling of the energy, we now impose $p_T > 200$~GeV
on the two leading jets.  The resultant $m_{jj}$ distribution is
plotted in Fig.\ref{fig:TeV_Axi_pT200_smeared}. The difference between
the {\OS} and the {\FL} cases is clearly visible even to the naked eye
and this shows that, even after smearing, the difference between the
two distributions is discernible.

It could be argued that, in making 
Fig.\ref{fig:TeV_Axi_pT200_smeared}, we have taken 
the liberty of finely binning the events. While this
demonstrates the difference between the two situations 
very well, such small bin sizes may not be achievable 
in practice. 
The study of Ref.~\cite{CDF_dijet_2008}, for example, 
uses a bin width of 0.1$m_{jj}$. 
To examine the effect of this reduced resolution, we redistribute
the events accordingly. The result is shown in 
Fig.\ref{fig:TeV_Axi_pT200_rebinned_smeared}.

\begin{figure}[!htbp]
\centering
\hspace*{-20pt}
\subfigure[]
{
\includegraphics[width=3.5in,height=3.0in]{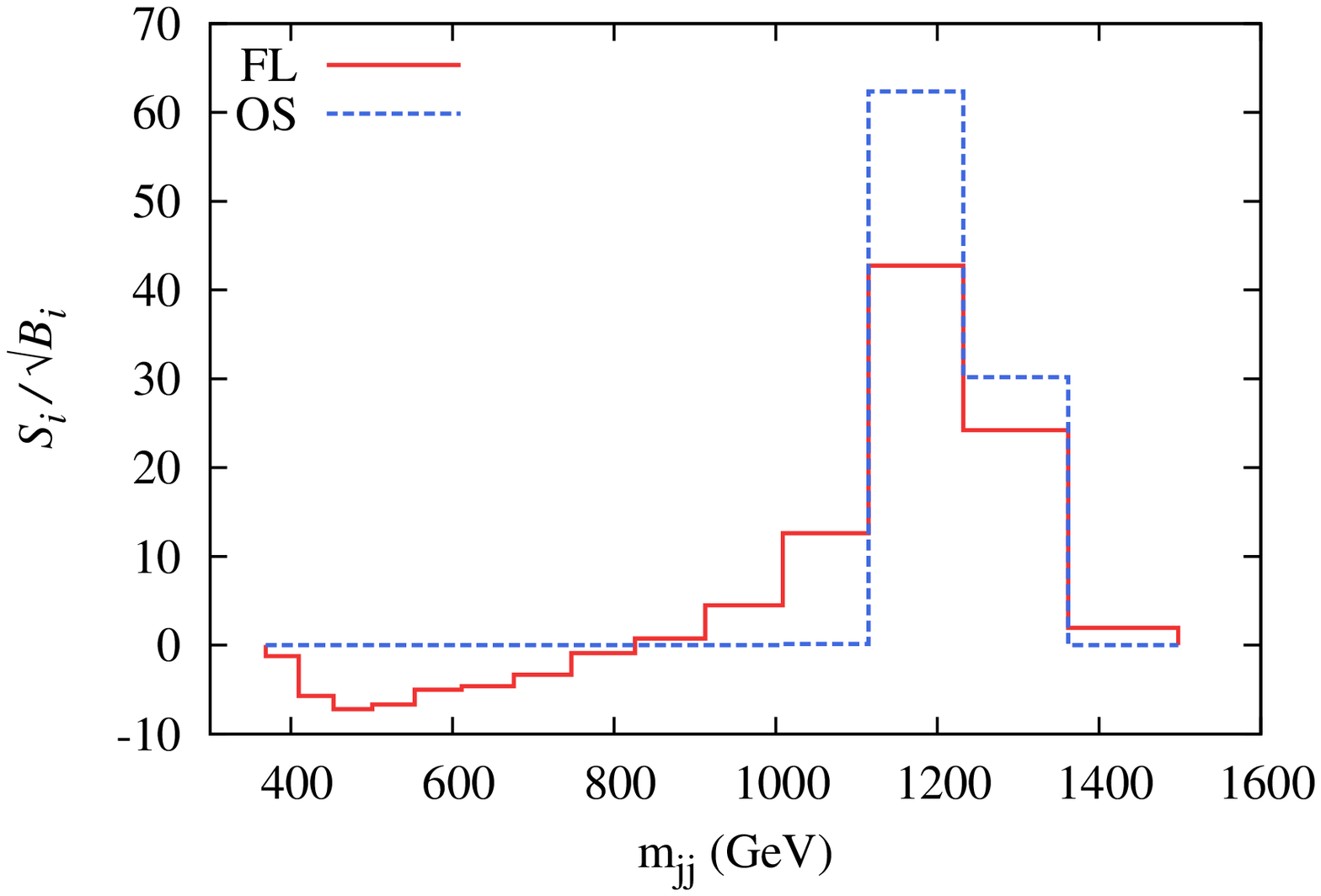}
\label{fig:rebinned_sigf_TeV}
}\hspace*{-10pt}
\subfigure[]
{
\includegraphics[width=3.5in,height=3.0in]{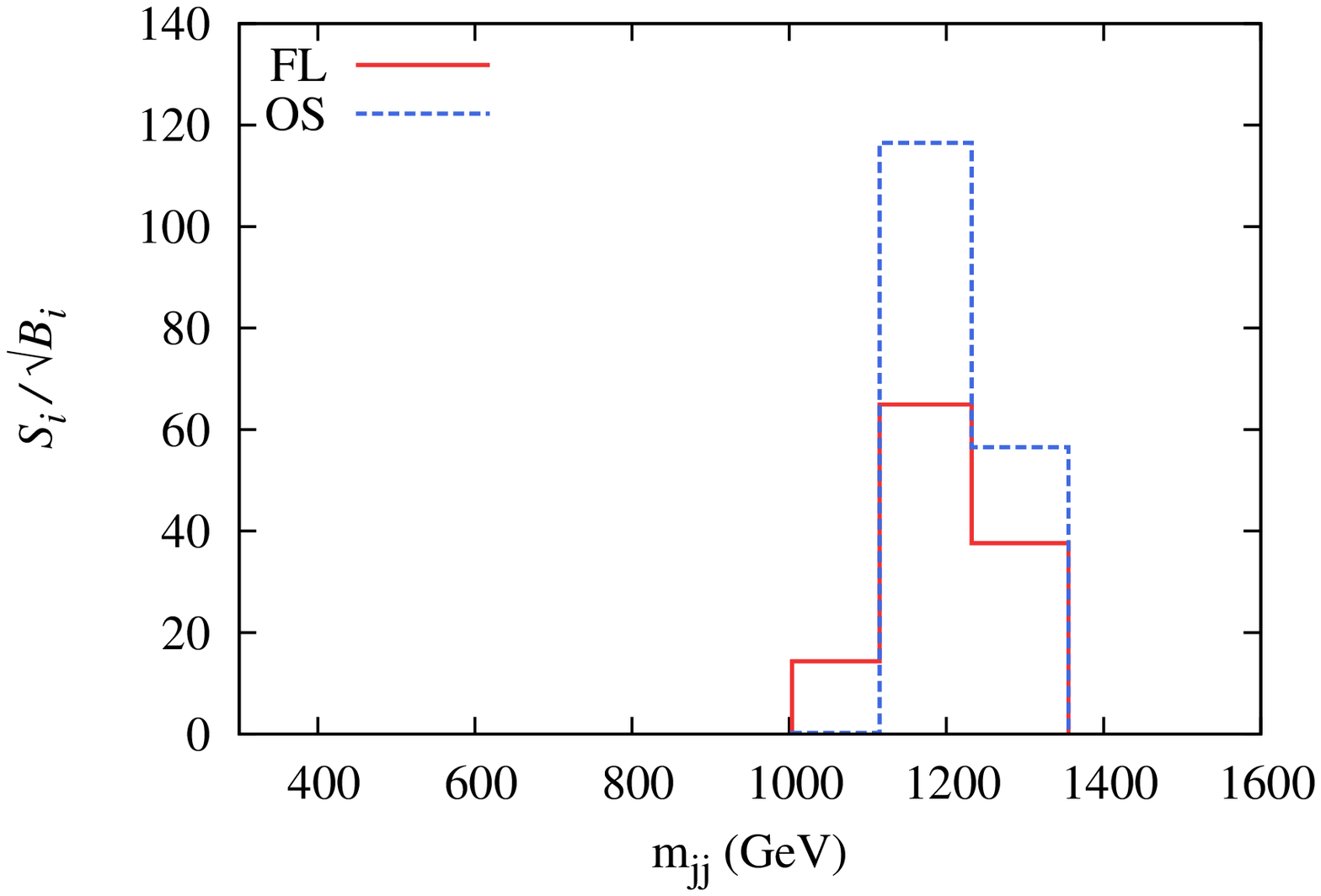}
\label{fig:rebinned_sigf_withcuts_TeV}
}
\caption{\em Binwise significance expected, at the Tevatron,
for an axigluon of mass $1.2$ {\tev} and bin-widths of $m_{jj}/10$. The two 
panels correspond to differing requirements on the two leading 
jets {\em (a)} $p_T \geq 200$ {\gev}, and {\em (b)} $p_T \geq 500$ {\gev}, $|y| < 0.7$.}
\label{fig:rebinned_significance_TeV}
\end{figure}

Understandably, the difference between the {\FL} and the {\OS} approximation
is reduced to an extent. Nonetheless, a significant difference between 
the two does persist as it does with the SM.
At this point, one must ask how significant
the difference is statistically.  To this end, we calculate the
binwise significance ${\cal S}_i$ defined as 
\[
{\cal S}_i \equiv S_i /\sqrt{B_i} \ ,
\]
where $S_i \, (B_i)$ are the number of signal (background)
events in a given bin\footnote{With the rebinning, the number of
events in each bin is large enough for the Poisson-distributed
number of events to be well-approximated by Gaussians.}.
The results, plotted in Fig.\ref{fig:rebinned_sigf_TeV}, 
bring out two facts. First, near the resonance, the significance 
obtained is much larger for the {\OS} case.  
Larger expectations of significance imply greater
sensitivity and allow the model to be ruled out at higher confidence
levels.  However, in this case, the greater significance is but an
artefact of the narrow-width approximation which, as demonstrated by
Fig.\ref{fig:TeV_Axi_pT200_smeared}, is clearly not applicable in this
case.  Second, away from the resonance, cross-sections are suppressed
and are {\em lower} than the SM prediction if one considers the full
({\FL}) amplitude\footnote{This suppression was 
not visible in Fig.\ref{fig:TeV_Axi_pT200_smeared} simply 
on account of the scale of the plots.}.
This is an effect of the interference of the
multiple axigluon-mediated contributions (including 
$t$-- and $u$--channel diagrams) with the SM amplitudes. 
In fact, with $p_T^{min}$ = 200 GeV, the total cross-section is slightly lower
than the corresponding SM prediction. While the dominant new physics 
contributions pertain to $q_i \bar q_i \to q_j \bar q_j$, even subdominant 
processes such as $q_i q_j \to q_i q_j$ are suppressed 
on account of the destructive interferences engendered 
by the $t$-channel axigluon exchange contributions.
Understandably, no such suppression exists in the
{\OS} approximation.

It is interesting to contemplate the impact of kinematic 
cuts on this difference. It seems plausible that placing a 
strong cut on the jet rapidities would not only
eliminate a larger fraction of the SM background, but also 
significantly reduce the contribution from the interference of 
the axigluon amplitudes with the $t$-channel gluon-exchange ones. 
In effect, the imposition of stronger requirements on the jet 
transverse momenta would be expected to push
the new physics contribution closer to the {\OS} approximation.
In Fig.\ref{fig:rebinned_sigf_withcuts_TeV}, we show the results for the 
significance on demanding $p_T^{\rm min} = $ 500 GeV and $|y| < 0.7$. 
Comparing it to Fig.\ref{fig:rebinned_sigf_TeV}, the increase in significance 
levels is obvious. What is even more striking is that the increase 
in significance is more pronounced for the {\OS} approximation than it is 
for the full ({\FL}) calculation. In other words, even the imposition of 
such strong cuts does not validate the {\OS} approximation, and the 
large sensitivity is partly illusory. The absence 
of a dip in the differential cross section (as compared to 
Fig.\ref{fig:rebinned_sigf_TeV}) is understandable as the corresponding 
$m_{jj}$ bins have been eliminated by the imposition of the 
strong $p_T$ cut.

\begin{figure}[!htbp]
\begin{center}
\includegraphics[width=4.0in,height=3.5in]{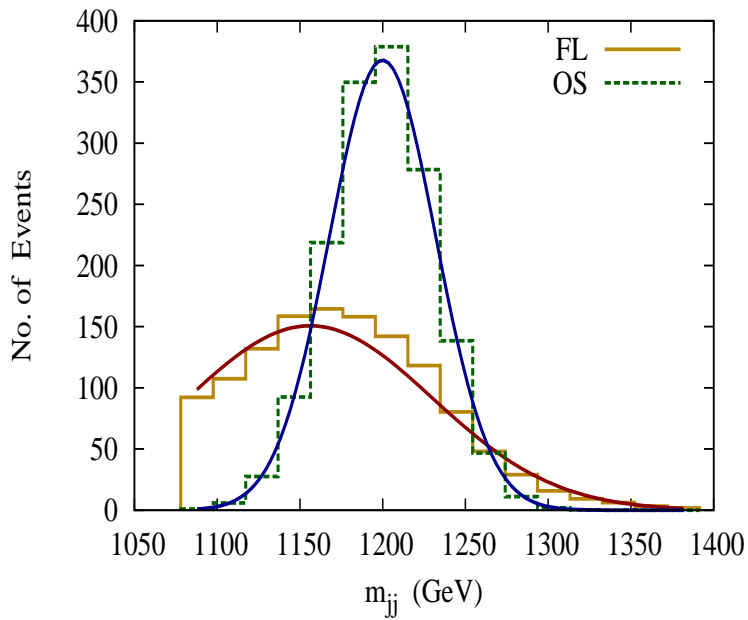}
\end{center}
\vspace*{-30pt}
\caption{\em The excess at the Tevatron 
(assuming \MA = $1200$ {\gev} and with $p_T^{\rm min} = 200$ {\gev}) 
after subtracting the SM background for the two scenarios  
{\em {\FL}} and {\em {\OS}}. The curves depict the result of 
the fitting when a Gaussian distribution is assumed.}
\label{fig:fits_TeV_pT200}
\end{figure}

Further, if one considers the background subtracted excess (as shown in 
Fig.~\ref{fig:fits_TeV_pT200}), one finds that, in case of {\OS},
the excess may be approximated by a Gaussian distribution.
On the other hand,  the {\FL} distribution is asymmetric and is not 
amenable to fitting by a simple functional form. Quite naturally then, 
if one were to attempt an extraction of parameters such as the mass and width, 
{\FL} and {\OS} would give rise to different values. In fact, this is borne out 
quite clearly by Fig.~\ref{fig:fits_TeV_pT200}. Not only is
the position of the peak visibly shifted in case of {\FL}, 
the shape too is sufficiently skewed to disallow an agreement with an 
{\OS} Monte Carlo.

\begin{figure}[!htbp]
\begin{center}
\includegraphics[width=4.0in,height=3.5in]{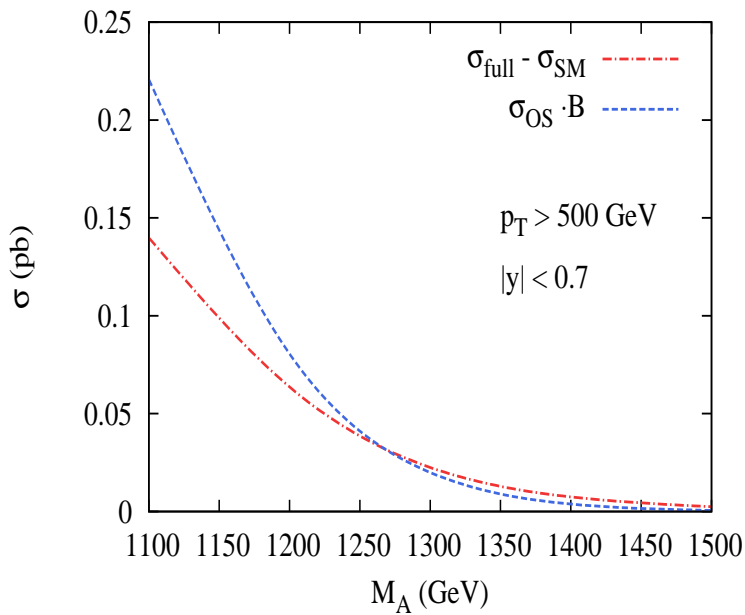}
\vspace*{-20pt}
\caption{\em The excess dijet cross section at the Tevatron as a function of 
the axigluon mass once each of the two leading jets have been 
required to satisfy $p_T > 500$ {\gev} and $|y| < 0.7$.}
\label{fig:excess_TeV_Axi}
\end{center}
\end{figure}

While the above discussion involved the invariant mass distribution which is 
the crucial observable in any resonance search, it is possible to 
appreciate the difference between {\OS} and {\FL} even by considering 
just the deviation in the total cross-sections, albeit with strong 
kinematic cuts. In Fig.\ref{fig:excess_TeV_Axi}
we show the deviation from the SM dijet cross-section as a function of \MA.
It is clear that considering simply on-shell production results in 
over-estimation of the signal for most cases that are within the 
kinematic reach of the Tevatron.
Again, this is a consequence of the multiple interference terms involving 
gluon/axigluon mediated $s$-, $t$- and $u$-channel 
amplitudes. In fact, were one to relax the cuts (for example, imposing 
only $p_T^{min} = $ 200 GeV with no restriction on rapidity), 
the effect of the interference would drive the total cross-section 
below the SM cross-section. It should be noted though that claiming 
a discovery (or otherwise) based on total rates alone is fraught with 
danger as this observable is particularly sensitive to higher order 
corrections, uncertainty in parton distributions etc. 

\begin{figure}[!htbp]
\begin{center}
\includegraphics[width=4.0in,height=3.5in]{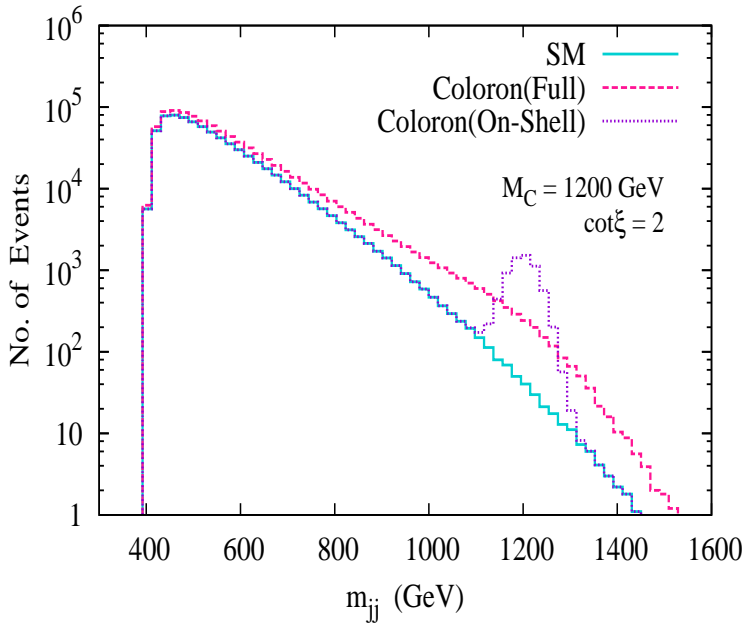}
\vspace*{-20pt}
\caption{\em The dijet invariant mass distribution, at the Tevatron, 
in the presence of a coloron with \MC = $1200$ {\gev} and $\cot\xi = 2$
(this translates to \GC = $408.5$ {\gev}).
Each of the two leading jets is required to have  $p_T > 200$ {\gev}
and an integrated luminosity of $10$ fb$^{-1}$ has been assumed.}
\label{fig:TeV_Col_pT200_smeared}
\end{center}
\end{figure}

The case of the coloron is even more curious.  The coupling strength
is $g_s \cot\xi$.  With $ 1 \leq \cot \xi \leq 4$, the widths could
easily be much larger than that for an axigluon of similar mass
($\sim$ 30\% of the mass with $\cot \xi$ of just 2). As a result, it
would be very difficult, if not impossible, to identify the excess as
a resonance (see Fig.\ref{fig:TeV_Col_pT200_smeared}).  Although it
might be argued that the enhancement in rates in the high $m_{jj}$
region\footnote{In fact, one of the motivations behind the proposal of
  the flavor-universal coloron model was the excess reported by the
  CDF collaboration in the tail of the $E_T$-spectrum for inclusive
  jet production~\cite{CDF_jet_excess}.  However, when more data was
  analysed, this discrepancy disappeared~\cite{CDF_jet_noexcess}.},
in itself, would constitute a smoking gun signal, note that the task
is not as straightforward. The wide region (in the $m_{jj}$ spectrum)
of the coloron's influence means that virtually no part of the
spectrum can be termed to be essentially SM-like, thereby changing the
entire nature of the fit algorithm. Thus, a much more sophisticated
algorithm, including a higher order calculation of the SM dijet
spectrum, would be necessitated; in particular, such corrections can,
and do, change the shape of the spectrum.

\subsection{At the LHC}
At the LHC, the pitch is queered by the low fluxes for antiquarks 
in the initial state. Due to this, the contribution from the 
$s$-channel amplitude suffers a reduction as compared 
to the contribution from the $t$- and $u$-channel amplitudes 
which may have two quarks in the initial state. 
Consequently, the {\OS} approximation is less likely 
be valid here as compared to the case of the Tevatron. 
For the SM background, a fit analogous to that for the Tevatron 
can be made~\cite{CMS_dijet_2011}, of course with differing parameters. 
Once again, a Gaussian fit works almost as well.

\begin{figure}[!htbp]
\centering
\hspace*{-20pt}
\subfigure[]
{
\includegraphics[width=3.5in,height=3.0in]{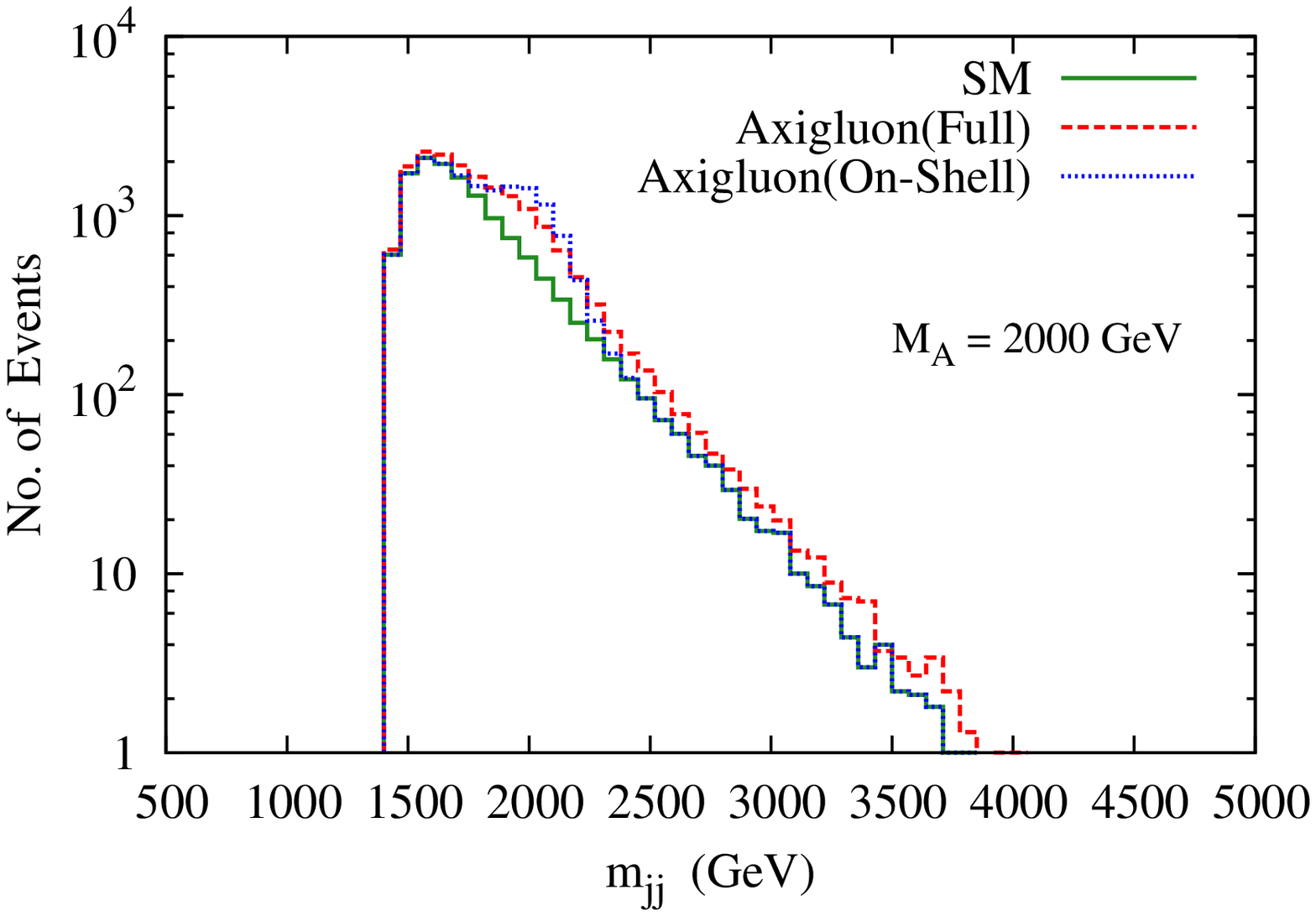}
\label{fig:LHC_Axi_pT700_y0.7_smeared}
}\hspace*{-10pt}
\subfigure[]
{
\includegraphics[width=3.5in,height=3.0in]{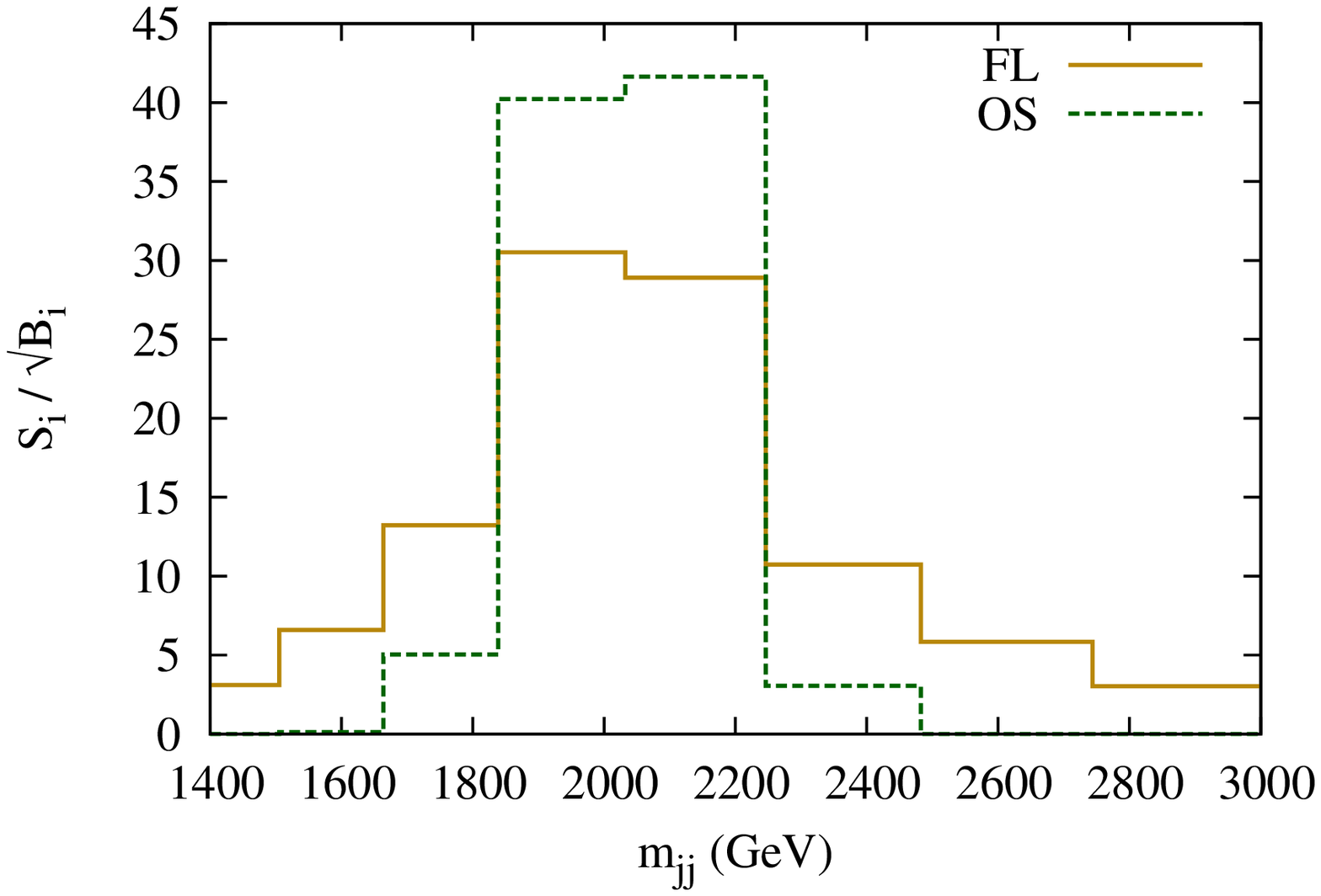}
\label{fig:rebinned_sigf_withcuts_LHC}
}
\caption{\em {\em (a)}The dijet invariant mass spectrum at the LHC 
corresponding to \MA = $2.0$ {\tev} \, \rm (\GA = 160.2 {\gev}). An integrated luminosity of 
$10$ fb$^{-1}$ has been assumed and a restriction of 
$p_T > 700$ {\gev} and $|y| < 0.7$ has been imposed 
on each of the two highest $p_T$ jets. {\em (b)} 
Expected binwise significance for the same case.}
\label{fig:LHC_Axi_withcuts}
\end{figure}

With the mass reach of the LHC being greater, we consider \MA = 2.0
TeV (\GA = 160.2 GeV), and, 
to enhance the signal to background ratio, we impose cuts
much stricter than those for the Tevatron, namely $p_T^{min} = $ 700
GeV and $|y| <$ 0.7.  For the energy resolution, we assume $\delta
E_T/E_T = 70\%/\sqrt{E_T ({\rm GeV})} \oplus
8\%$~\cite{LHC_resolutions}.  After smearing, a resonance-like
structure seems apparent for the {\OS} case, though no such claim can
be made for {\FL} (Fig.~\ref{fig:LHC_Axi_pT700_y0.7_smeared}).
However, once larger (and experimentally more reasonable) 
bin-widths\footnote{Though we use bin-widths of 0.1$m_{jj}$ here, 
at ATLAS, for example, a resolution of 0.05$m_{jj}$ 
is possible~\cite{giacomo}.}
are considered, the difference is expected to reduce.  Note that part
of this effect is offset by the increased statistics in each bin.  As
Fig.~\ref{fig:rebinned_sigf_withcuts_LHC} shows, even with larger
bin-widths, {\OS} yields higher values for significance close to the
resonance, whereas {\FL} gives rise to significant deviations away
from the resonance.

\begin{figure}[!htbp]
\begin{center}
\includegraphics[width=4.0in,height=3.0in]{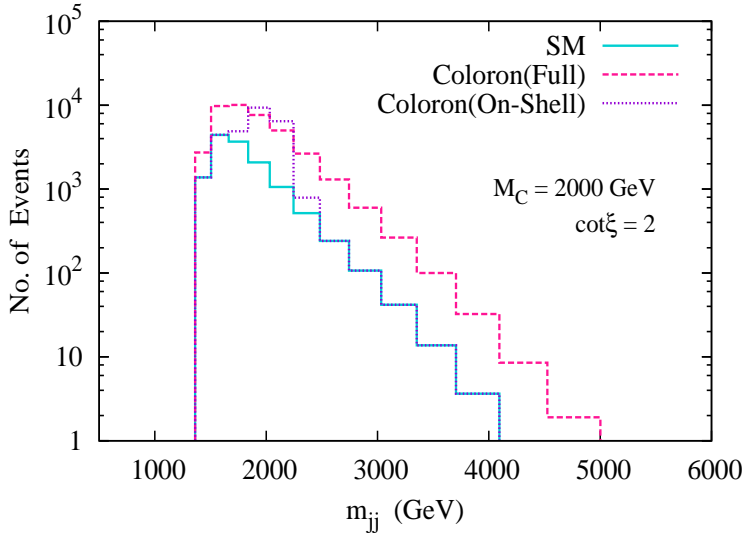}
\vspace*{-20pt}
\caption{\em The dijet invariant mass spectrum at the LHC corresponding 
to \MC = $2.0$ {\tev} and $\cot\xi = 2$ \rm (\GC = 645.7 {\gev}). An integrated luminosity of 
$10$ fb$^{-1}$ has been
assumed and a restriction of $p_T > 700$ {\gev} and $|y| < 0.7$ has been imposed 
on each of the two highest $p_T$ jets.}
\label{fig:LHC_Col_pT700_y0.7_rebinned_smeared}
\end{center}
\end{figure}

Finally, we present the comparison between the invariant mass distributions 
for a coloron with \MC = 2.0 TeV and $\cot\xi = 2$ (\GC = 645.7 GeV)
in Fig.\ref{fig:LHC_Col_pT700_y0.7_rebinned_smeared}. 
As with the case of the Tevatron, the {\OS} approximation is 
demonstrably a very poor one and almost the entire dijet mass spectrum 
gets modified. 

\subsubsection{Other Observables}
We have, until now, concentrated only on the invariant mass distribution 
as a discriminator, not only between the SM and new physics, but 
also amongst various forms of the latter. With the resolving power 
decreasing as the natural width increases, the case for other variables 
such as angular distributions becomes progressively stronger.  
For example, it has been argued~\cite{Haisch:2011up} that the 
latter could be used to distinguish between a broad resonance 
and contact interactions.

\begin{figure}[!htbp]
\centering
\hspace*{-20pt}
\subfigure[]
{
\includegraphics[width=3.5in,height=3.0in]{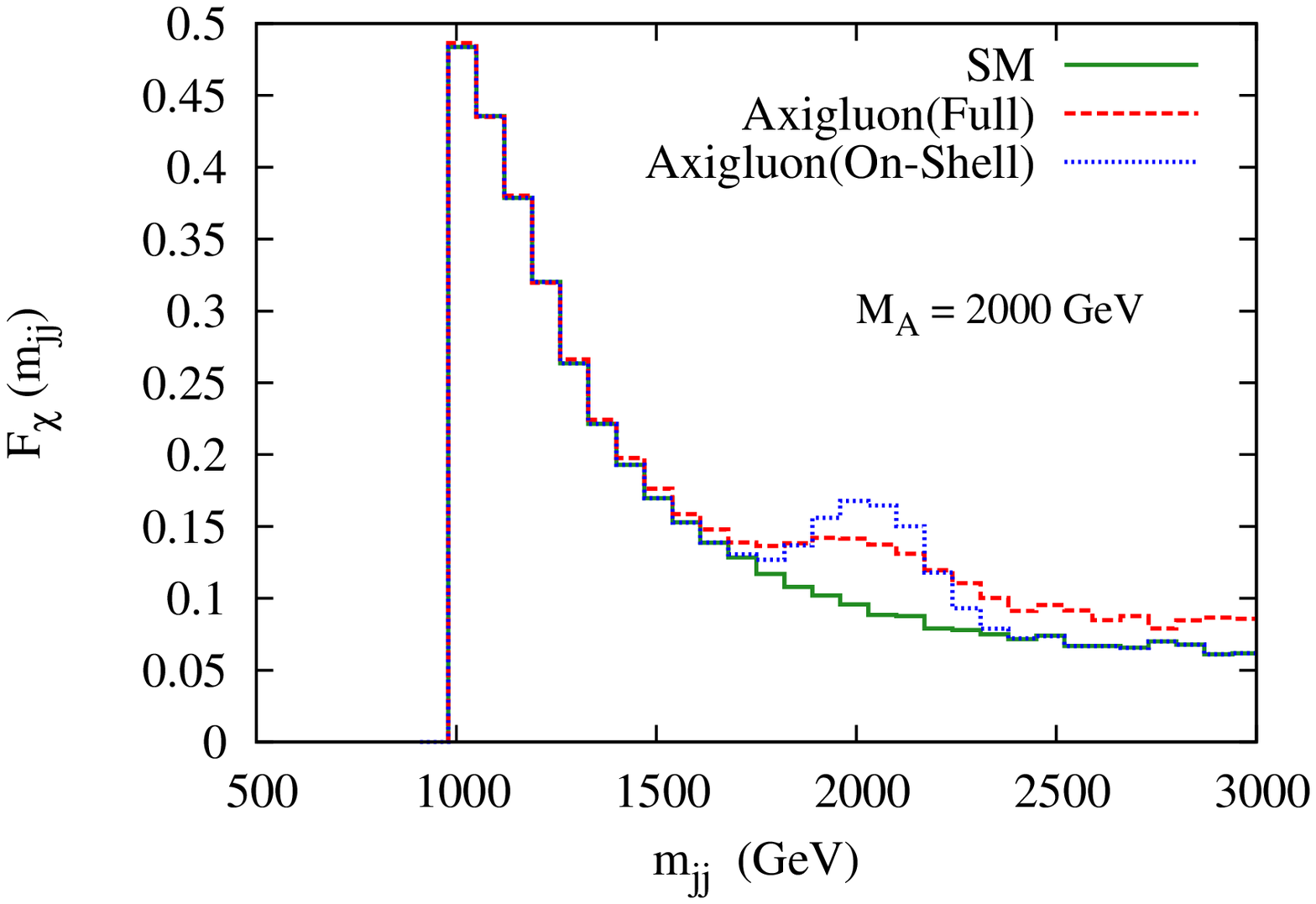}
\label{fig:Fchi}
}\hspace*{-10pt}
\subfigure[]
{
\includegraphics[width=3.5in,height=3.0in]{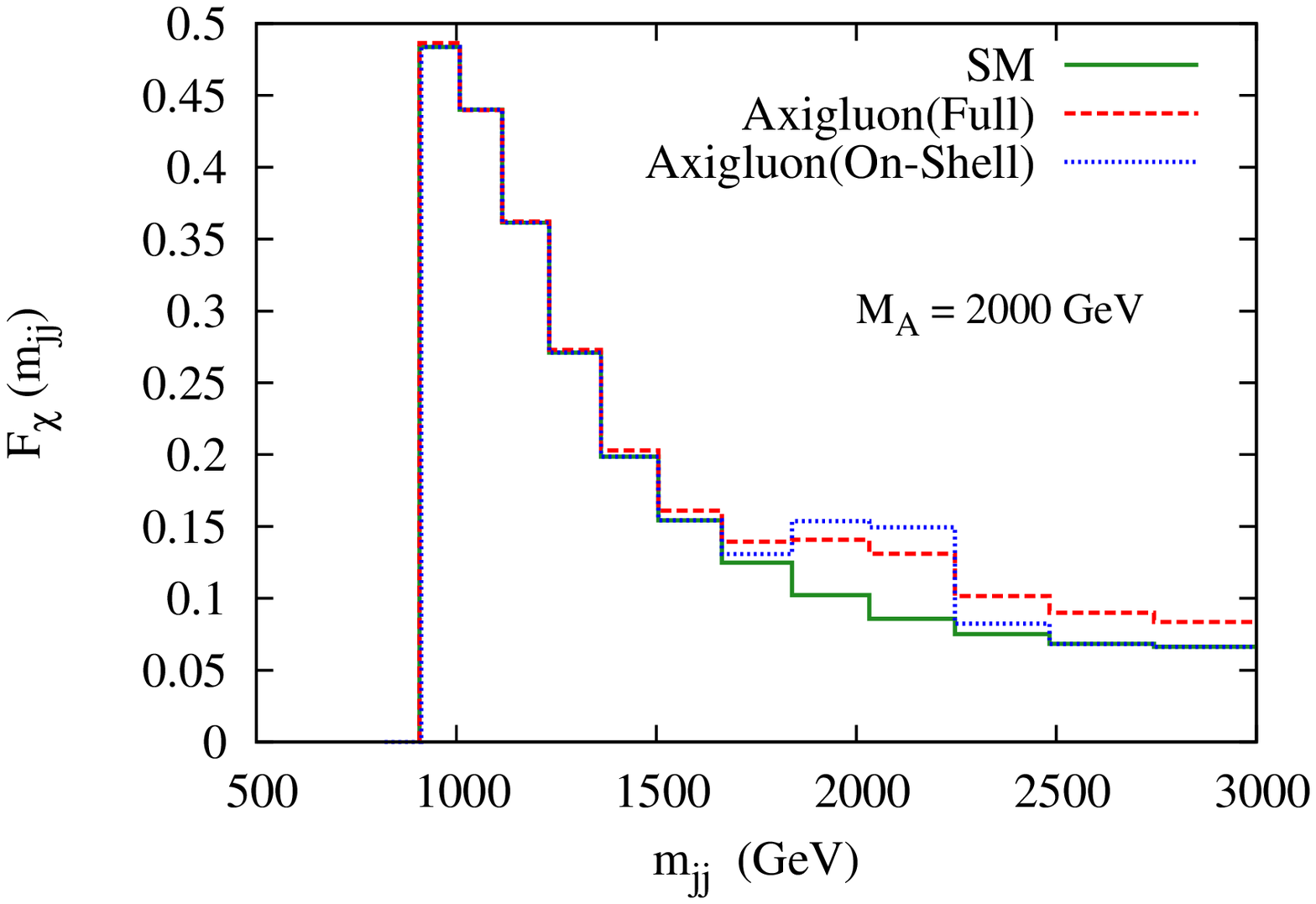}
\label{fig:Fchi_rebinned}
}
\caption{\em The $F_{\chi}(m_{jj})$ distribution at the LHC for 
{\MA}{\em = 2.0 {\tev}}  
with $p_T^{min}${\em = 500 {\gev}}. In (b), the bin-width 
taken to be {\em 0.1}$m_{jj}$.}
\label{fig:Fchi_analysis}
\end{figure}

In this context, we re-examine a variable that has been used 
by the ATLAS collaboration~\cite{ATLAS_Fchi} to obtain limits 
on quark contact interactions. In essence, they compare the number 
of events where both jets satisfy $|y^*| < 0.6$ (where $y^*$ 
is the rapidity in the partonic CM frame) with that where the 
extent of centrality is relaxed to  $|y^*| < 1.7$. To be precise, 
$F_{\chi}(m_{jj})$ is defined as~\cite{ATLAS_Fchi} 
\[
F_{\chi}^i(m_{jj}) 
=
\dfrac{N_{events}^i(|y^*| < 0.6)}{N_{events}^i(|y^*| < 1.7)},
\]
where, the superscript $i$ denotes the $i^{th}$ bin in the $m_{jj}$
distribution.  We compute the $F_{\chi}$ distributions for an axigluon
of mass 2000 GeV for each of the {\FL} and {\OS} cases. As suggested
by Ref.~\cite{ATLAS_Fchi}, the $F_{\chi}$ distribution is indeed
sensitive to mass dependent changes in production rates in the central
rapidity region. But that is not all. It is also sensitive to the
assumption of a narrow-width approximation. In other words, as can be
seen in Fig.~\ref{fig:Fchi} the shape of the distribution is markedly
different in the two cases\footnote{That the shape of our SM curve are
different from that in Ref.~\cite{ATLAS_Fchi} is attributable to
the strong $p_T$ cut imposed by us.}. The difference persists even 
when the bin-size is increased (Fig.\ref{fig:Fchi_rebinned}).

\section{Summary}
\label{sec:summary}
New physics scenarios often predict resonances in the dijet channel.
Very often, search strategies focus on setting limits on $\sigma \cdot
{\cal B}$, the product of the cross section for on-shell production
and the branching fraction into the dijet channel. While this strategy
is perfectly valid in many cases, its applicability is questionable in
the case of particles which have large widths due to either strong interactions 
or large masses. It has been argued in the literature~\cite{CDF_dijet_2008}, though, that
the natural width plays only a subservient role to detector resolution
effects, and, consequently, the former is oft neglected in
experimental strategies.

To examine this assertion in a qualitative manner, we considered two
models and, for each, performed a simple analysis comparing
the signal profile with and without the `narrow-width
approximation'. The dominant experimental effects (resolution) were
incorporated through a smearing of the energies of the final-state
particles\footnote{Indeed, we made a conservative choice of resolution
  parameters so as to enhance the experimental effects.}.  We do find
that the profiles can be substantially different in the two cases.
For example, in the axigluon case, the natural width broadens the
invariant mass distribution to a significant extent over and above the
broadening due to resolution. 
A further complication arises from the fact that, in addition to the $s$-channel
diagrams, we also have $t$-- and $u$--channel contributions. While
these can be safely neglected for a narrow resonance, their importance
increases as the width becomes larger. In addition, channels such as
$u \bar d \to u \bar d$ which do not proceed through $s$-channel
diagrams do receive contributions from a $t$-channel axigluon
exchange. In totality, it transpires that inclusion of such effects 
would lead to tangible {\em worsening} of the exclusion limits. 

At the LHC, these effects could be even more important.
The reason is not difficult to fathom.  At the LHC, the $\bar q$ is a sea-quark and,
consequently, its density is relatively small for high Bjorken-$x$,
the regime that producing a very heavy resonance would require. On the
other hand, $t$-channel axigluon contributions to $q q \to q q $
remain unsuppressed. Thus, a careful analysis is even more mandatory 
in that arena.

The signal to background ratio is, of course, dependent on the kinematical cuts. 
We have demonstrated that this dependance itself is quite different for the
{\OS} case as compared to the full calculation.  Thus, we feel that the
analyses should preferably be done with the inclusion of all diagrams
and, possibly retuning the selection cuts so as to obtain robust
exclusion limits. Moreover, while we have presented only tree-level results here, 
contributions from higher orders in perturbation theory would also have to be 
taken into account, both for the SM as well as the relevant new physics scenario. 

While the preceding discussion has largely focussed on the axigluon,
it is by no means the only example of a colored resonance with a large
width. Indeed, colorons generically have even larger widths. As we
have shown explicitly, considering even $\cot \xi = 2$ (where,
theoretically, $1 \leq \cot \xi \lesssim 4$) would render the 
width quite large and would almost totally obliterate a
resonance. Rather, a very broad excess would be visible. Such an
excess, in principle, could stand out once the background has been
subtracted. However, the very width makes it virtually impossible to
fit a functional form to the observations away from an excess, for,
indeed, the excess spans almost the entire available invariant mass
range. Hence, identification of the excess would require either a very
precise knowledge of the SM prediction and perhaps even new experimental techniques.

\begin{center}\begin{small}\textbf{ACKNOWLEDGEMENT}\end{small}\end{center}
We thank A. Harel for useful suggestions.
R.M.G. wishes to acknowledge support from the Department of Science and
Technology, India under Grant No. SR/S2/JCB-64/2007.
P.S. would like to thank CSIR, India for assistance under 
SRF Grant 09/045(0736)/2008-EMR-I.



\end{document}